# Real time reconstruction of the fast electron spectrum from high intensity laser plasma interaction using gamma counting technique

A. Zavorotnyi [a,1], A. Savel'ev [a,b]

[a] *Faculty of Physics, M.V. Lomonosov Moscow State University,*
*1, Leninskie gory, 119991, Moscow, Russia*

[b] *Lebedev Physical Institute of the Russian Academy of Sciences,*
*53 Leninskiy Prospekt, 119991, Moscow, Russia*
*E-mail*: akimzav@gmail.com

ABSTRACT: X-ray and gamma fluxes from the high intensity laser-plasma interaction are extremely short, well beyond temporal resolution of any detectors. If laser pulses come repetitively, the single photon counting technique allows to accumulate the photon spectra, however, its relation to the spectrum of the initial fast electron population in plasma is not straightforward. We present efficient and fast approach based on the Geant4 package that significantly reduces computer time needed to re-construct the high energy tail of electron spectrum from experimental data accounting for the pileup effect. Here, we first tabulate gamma spectrum from monoenergetic electron bunches of different energy for a given experimental setup, and then compose the simulated spectrum. To account for the pileups, we derive analytical formula to reverse the data. We also consider errors coming from the approximation of the initial electron spectrum by the sum of monoenergetic impacts, the finite range of the electron spectrum, etc. and give estimates on how to choose modelling parameters to minimize the approximation errors. Finally, we present an example of the experimental data treatment for the case of laser-solid interaction using 50 fs laser pulse with intensity above $10^{18}$ W/cm$^2$.

---

[1] Corresponding author.

# Contents

## 1. Introduction

In recent decades, advances have been made in the field of femtosecond laser technology, making it possible to achieve tera- and petawatt laser peak powers with pulse duration down to a few optical periods. In many ways, this was facilitated by research on nonlinear optics, the acceleration of charged particles, controlled thermonuclear fusion, and others [1] [2] [3] [4] [5] [6]. Focusing of high-power femtosecond laser radiation makes it possible to obtain peak intensity from $10^{18}$ to $10^{23}$ W/cm$^2$ and provides for the field ionization of target atoms already at the front of the laser pulse [7] [8]. The laser plasma formed during such an interaction is of great interest due to its unique properties, among which one of the most important is a nonthermal distribution of electrons: fast, or hot, electrons appear due to collisionless absorption of laser energy or electron's acceleration by the laser induced plasma field [9]. Mean energy of such electrons varies from tenths of keV to hundreds of MeV depending on the interaction regime [10] [11] [12] [13] [14] [15].

     Penetration of electrons deep into the target or their interaction with additional target-converter leads to generation of bremsstrahlung and line hard X-ray and gamma radiation [4] [5] [16]. Laser-plasma sources of this kind are widely used for temporally and spatially resolved X-ray studies, in nuclear photonics [17], medicine and biology etc. [1] [2] [3] [4]. Optimization of the laser-plasma interaction demands knowledge of the parameters of accelerated electrons. In some cases this can be done directly using magnetic spectrometers [18], layered detectors [19], [20], [21], [22], [23], etc. Alternatively, the bremsstrahlung spectrum obtained when electrons are scattered by a solid target can be used [24], [25], since the measured X-ray spectrum is determined by the initial spectrum of accelerated electrons. If the laser works in a

repetitive regime, a single event gamma detector (such as scintillator coupled with a PMT) is used with output pulse amplitude proportional to the energy absorbed by a scintillator [21]. Hence less than one gammas should be detected for each laser pulse to get the spectrum, and to avoid the pileups (simultaneous detection of a few quanta) the counting rate has to be much less than unity.

Quantitative approach involves analysis of the X-ray spectra obtained in the experiment using computer simulation. Most often, the Geant4 package [26] is used to establish the relationship between the X-ray (or gamma) and electrons spectra. Here the initial electron spectrum is varied (most often – its slope assuming exponential spectrum of electrons), and numerical calculation is performed for each set of parameters to get an X-ray spectrum that fits the measured one.

This approach has the crucial disadvantage. Since the typical electron spectrum is an exponential one, computer time needed to reproduce the high energy tail of an x-ray spectrum with enough statistics grows up exponentially if the considered X-ray energy interval is enlarged. Note, that for many applications this tail is the most important issue to be retrieved. The purpose of our work is to demonstrate the new approach to the experimental data processing, which makes it possible to retrieve electron spectrum slope and amplitude much faster. We used the pre-calculated response functions (obtained using Geant4 within the given experimental geometry) for a set of electron energies to fit the experimentally measured X-ray spectrum. Here the initial guess on the electron spectrum envelope is needed. This fitting procedure is very fast. Besides we derived analytical formula describing the pile-up effect in the case of bi-exponential electron distribution. Incorporating it into the fitting procedure we successfully restored the electron spectrum from the X-ray one obtained at the counting detector rate as high as 0.3 in a wide spectral range.

## 2. Problem formulation

Typical experimental setup is presented in Figure 1. Accelerated electrons of spectrum $n(E_e)$ (1) interact with all parts of the experimental setup (2), which includes a scintillating crystal (*). Inelastic interactions of both electrons thereselves and generated X- and gamma quanta with the crystal produce so-called events. Energies $E_x$ deposited by these events are recorded by the PMT-computer junction (3). It is crucial here to have only single event for each laser shot, otherwise, the pile-up comes into play. Repeating this cycle and accumulating events over many iterations, the spectrum $n'(E)$ is formed that contains both electron and photon impacts.

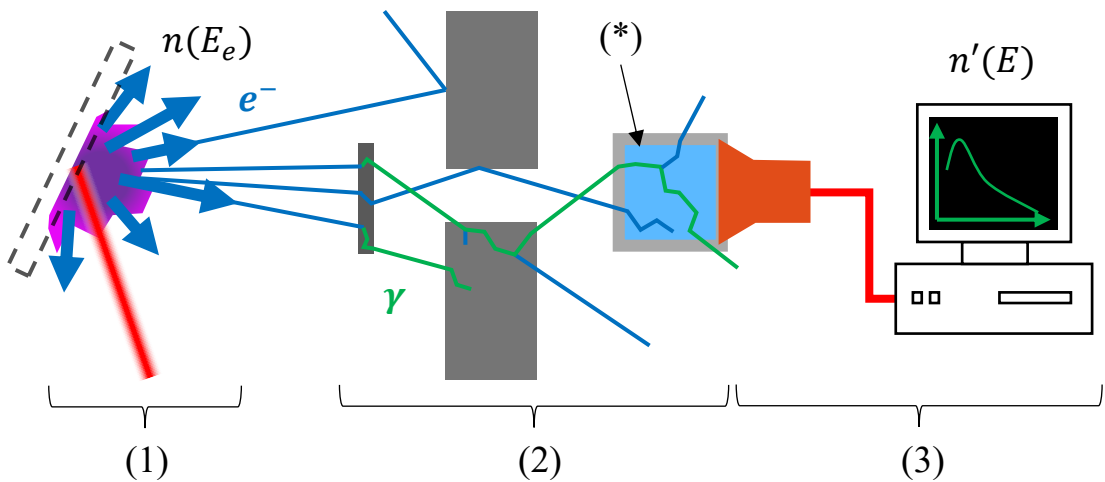

**Figure 1.** Typical experimental setup for the X-ray spectrum accumulation: interaction area (1); target converter (2) with the (typically) lead diaphragm and scintillator crystal (*); PMT, ADC and the computer (3).

When considering this technique, the problem of interpreting the experimentally measured spectrum arises. In the most unspecific way, it can be formulated as follows: «Let the spectrum $n'(E)$ be measured by the detector. What was the spectrum of accelerated electrons $n(E_e)$?». The physical conditions of the experiment – the characteristic dimensions and energies – are such that it is rational to consider the participating particles as material points. In addition, two-, three-, and more particle interactions can be neglected, from which it follows that it is possible to consider all participating particles separately. This automatically implies the linearity of the transformation $n(E_e) \rightarrow n'(E)$.

In the following we will discuss measurements fulfilled with the solid laser plasma interaction. In this case the initial fast electron spectrum may be in first approximation assumed as the sum of two exponents:

$$n(E_e) = Ae^{-E_e/T_1} + Be^{-E_e/T_2}. \qquad (1)$$

Here the first term corresponds to the "low" energy electrons and has higher amplitude $(A \gg B)$, while the most interesting part is the second term. As it follows from numerous experimental data [19] [20] [21] the high energy tail of the secondary spectrum $n'(E)$ can be also presented in the exponential form:

$$n'(E) \propto e^{-E/T'}. \qquad (2)$$

It is natural to assume that this high energy tail is due to the second term, hence the key goal of our study is to establish procedures to get estimation on the $T_2$ knowing $T'$.

To simulate the measurement procedure described above we choose the software package Geant4 [27] for the numerical Monte-Carlo simulation on the track-level. It consists in simulating single particles' tracks with the initial parameters (the electron spectrum $n(E_e)$ in our case) defining distribution of the corresponding particle's initial properties. The output values of the simulation come from the numerical detector – an entity modeling the real detector. In our case, an event recorded by the numerical detector is the energy $E$ absorbed within the scintillator's volume as a result of either inelastic interaction of primary electrons, secondary electrons, or X- and gamma quanta. The Photomultiplier tube is excluded from the simulation by the reasonable assumption that the number of photons emitted by the scintillator is linearly proportional to the energy absorbed by the scintillator.

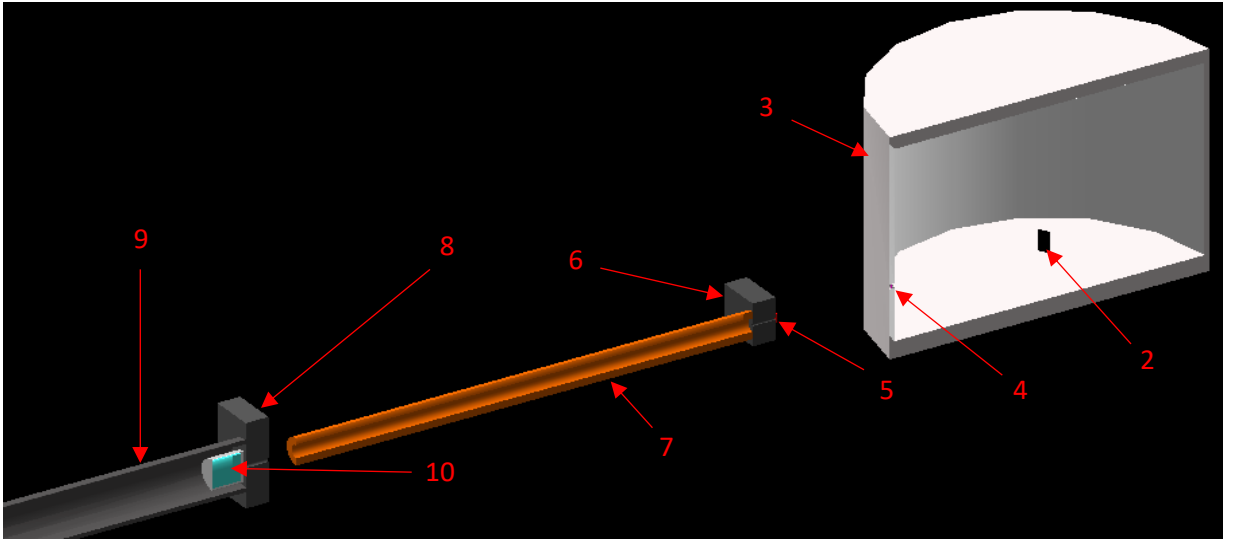

**Figure 2.** Experimental setup, designed in Geant4, including target (2), vacuum chamber (3), outlet window (4), filter (5), collimators (6-8), shielding (9), scintillation crystal (10).

Naturally, we could run our simulations for different values of the parameters $A, B, T_1, T_2$ and get the required dependence $T'(T_2)$. However, there are two key problems that such the modeling would face:

1. The problem of high-energy simulation: due to the fact that $A \gg B$, it is difficult to collect statistics for the high energy events;
2. The problem of the pileup effect [25]: each high energy particle detected by the scintillator may be accompanied by a few secondary low energy ones, if the temporal resolution of the detector is less than time delays between the events. Hence the detector would assume this as a detection of a single particle with overestimated energy.

Let's consider them in more details.

## 2.1. The problem of high-energy simulation

The time required to collect sufficient statistical weight for a given energy $E$ increases exponentially with increasing $E$, if the spectrum $n'(E)$ decreases exponentially. For example, it takes 10 hours with the server based on the Intel Xeon X5675 processor to collect at least 10 events in 200 keV-wide bins up to 1 MeV if $T_2$=1500 keV, and 27 hours if the upper threshold is 2 MeV, 74 hours for 3 MeV, 201 hours for 4 MeV, 546 hours for 5 MeV, and so on. This is a significant problem, fundamentally limiting the feasible range of $E$ for modeling.

This problem is further aggravated by the fact that the approach outlined tacitly assumes that the calculation is carried out a few times to get dependence $T'(T_2)$ and other parameters.

Thus, the "direct" approach (i.e., simulation of the observed spectrum for a set of initial spectra) turns out to be inapplicable in practice for a limited computational time.

## 2.2. The pileup effect

Two or more gamma quanta or electrons may enter the scintillator simultaneously (or within the temporal resolution of the detector), with the sum of their energies recorded. It is clear that the observed spectrum might change qualitatively [25] : an overestimation of the slope $T_2$ by 15 -70% occurs depending on the detector properties. That is why the detector counting rate $\mu$ should be kept well smaller than 1, increasing experimental time needed to get the high energy spectrum with acceptable statistics. This problem becomes even worse, if the bi-exponential electron distribution is assumed, since each high energy particle may be accompanied by a few low energy ones, increasing the distortion of the higher-energy tail. Since typical value of $\mu$ amounts to 0.1…0.3, the influence of pileup is non-negligible, so it is necessary to find out a method of transforming the Geant4-simulated (pileup-free) spectrum into the spectrum with the same counting rate as in experiment, thus taking the pileup effect into consideration. Note that direct (Monte-Carlo) pileup simulation is not applicable, as in case of the bi-exponential spectrum the time required by such simulation increases exponentially with the linear increase of the monitored energy range.

As such, a derivation of an explicit, specific functional describing the $n'(E) \to n''(E)$ transformation is required.

This paper presents algorithms, procedures and results showing how to assess the initial electron spectrum from the experimental one overcoming the problem of exponential growth of simulation time for high energies and considering the pileup effect using a transformation of this kind.

## 3. Problem solving

### 3.1. Delta-function approximation technique

First, we will neglect the contribution of the pileup effect, and consider the problem of high energy simulation. Let us approximate a spectrum $n(E_e)$ as the sum: $n(E_e) = \sum_{i=1}^{M} a_i \delta(E_e - E_{ei})$, where $a_i$ and $E_{ei}$ are known values. Hence,

$$n'(E) = \sum_{i=1}^{M} a_i n'_i(E), \tag{3}$$

where $n'_i(E)$ are the spectra detected with the initial electron spectrum $n_i(E_e) = \delta(E_e - E_{ei})$. Accordingly, having calculated a set of values $n'_i(E)$ through the simulation, we can get an estimate for $n'(E)$, using the formula above.

Visualization of the method is presented in Figure 3. It consists of the following steps:

I. Preprocessing with a given experimental arrangement:

1. Define a set of energies $\{E_{ei}\}$.
2. For each $E_{ei}$ in the set, use Geant4 simulation with a monoenergetic initial spectrum $n_i(E_e) = \delta(E_e - E_{ei})$, acquiring $n'_i(E)$. Once done, a set $\{n'_i(E)\}$ is obtained and stored.

II. Spectrum calculation for a given $n(E_e)$

1. For an assumed initial electron spectrum $n(E_e)$ calculate corresponding set of coefficients $a_i = \int_{E_{ei-1}}^{E_{ei}} n(E)\, dE$.
2. Calculate the output spectrum $n'(E) = \sum_i (a_i * n'_i(E))$.

Note, that once the set $\{n'_i(E)\}$ has been calculated at the first stage, it can be used repeatedly for different set of parameters describing the spectrum and different experiments provided the experimental setup remains unchanged. This reduces the computation time tremendously, since the second stage assumes simple algebraic manipulations are used, while additional Geant4 simulations are not needed here.

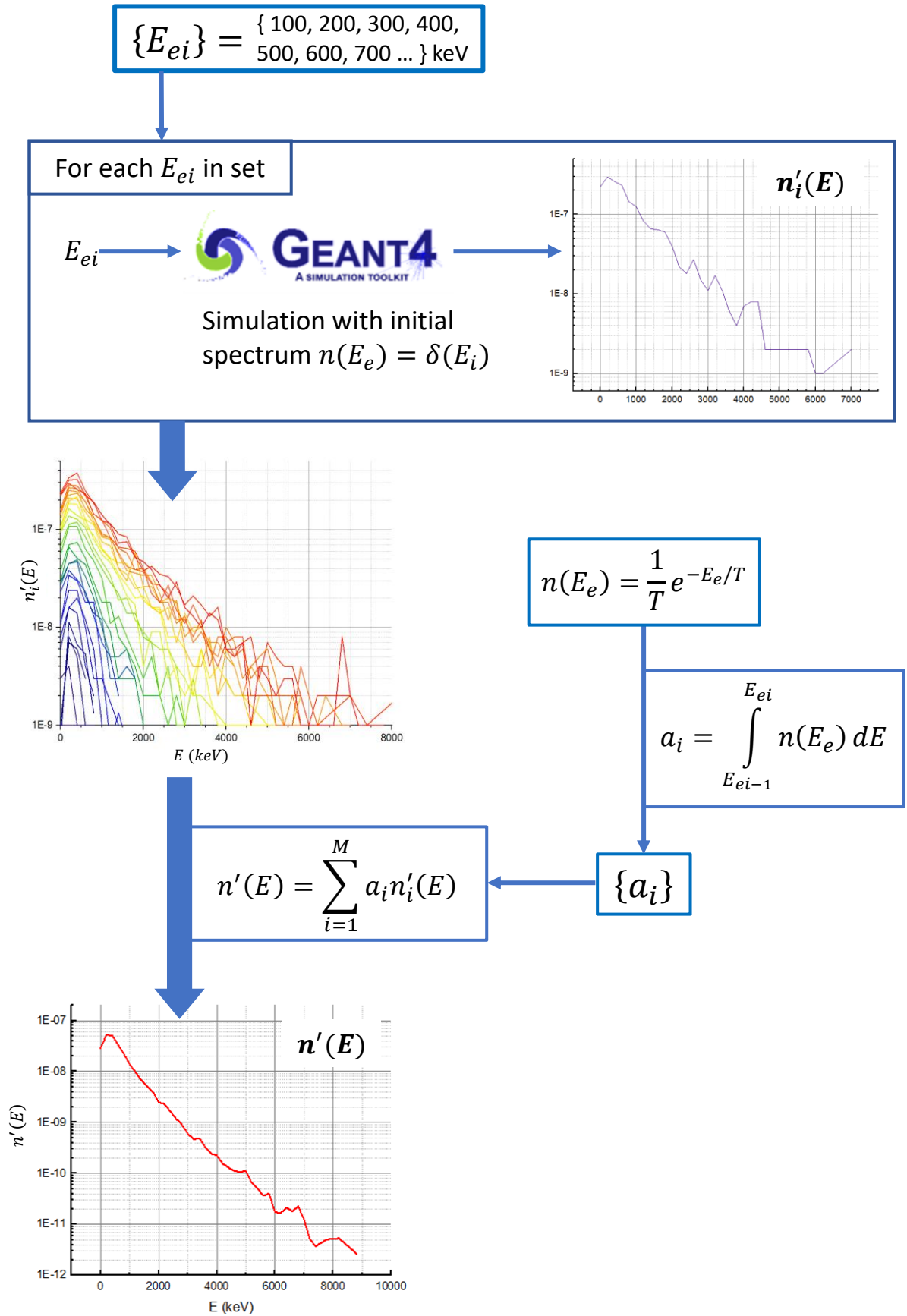

**Figure 3.** Graphical representation of the method's algorithm.

Our method introduces two types of errors: (i) due to the replacement of a continuous function by a discrete one and (ii) due to a finite range, covered by the $\{E_{ei}\}$ set. Let's consider the importance of those errors.

It can be shown that if the following condition is true:

$$\left|\left(\frac{\partial n_C(E_e)}{\partial E_e} \cdot (E_{ei} - E_{ei-1})\right) / n_C(E_e)\right| \ll 1 \quad (4)$$

for each $E_{ei} \in \{E_{ei}\}$, where $n_C(E_e)$ – electron spectrum in its continuous form, then, assuming $n_D(E_e) = \sum_{i=1}^{M} a_i \delta(E_e - E_{ei})$ is the discrete representation of the initial spectrum, the difference $\Delta n'_{CD}(E) = n'_C(E) - n'_D(E)$ between the simulated spectra can be written as (see Appendix 1):

$$\Delta n'_{CD}(E) \cong \frac{1}{2} \sum_{i=1}^{\infty} \left.\frac{\partial p(E_e, E)}{\partial E_e}\right|_{E_e = E_{ei}} \cdot n_C(E_{ei}) \cdot (E_{ei} - E_{ei-1})^2, \quad (5)$$

Here transformation function $p(E_e, E)$ is defined from $n'(E) = \int_0^\infty p(E_e, E) n(E_e) dE_e$. The aforementioned condition, in case of $n(E_e) = \frac{1}{T} e^{-E_e/T}$ can be simplified as:

$$(E_{ei} - E_{ei-1})/T \ll 1. \quad (6)$$

It means that, if the neighboring energies from the set $\{E_{ei}\}$ are close enough to each other, then the error associated with the fact that a continuous function is replaced by a sum of delta functions grows not faster than the square of the distance between them, i.e. proportionally to $(E_{ei} - E_{ei-1})^2$.

It can be shown also, that if: (i) the value of $n'_i(E)$ at a given $E$ with increase of $E_{ei}$ grows no faster than linearly, and (ii) function $n'(E)$ has a decreasing exponential asymptotic, the error at the given $E$ decreases exponentially with the number of peaks added to the end of the set $\{E_{ei}\}$, if such peaks are equidistant (see Appendix 2).

### 3.2. Pileup reversing method

The analytical derivation of the mathematical transformation $n'(E) \to n''(E)$ of the observed spectrum $n'(E)$ into the spectrum $n''(E)$ observed in the presence of pileups is presented in the Appendix 3. In the case of bi-exponential spectrum the probability density function $P(E)$ to detect an event with a given energy $E$, given average number of registration events $\mu$ reads as:

$$P(E) = \left(p_1 - p_2 * 2A_2 \frac{T_1 T_2}{T_2 - T_1}\right) * A_1 e^{-\frac{E}{T_1}} + p_2 * A_1^2 * E e^{-\frac{E}{T_1}} + \\ + \left(p_1 - p_2 * 2A_1 \frac{T_1 T_2}{T_1 - T_2}\right) * A_2 e^{-\frac{E}{T_2}} + p_2 * A_2^2 * E e^{-\frac{E}{T_2}}, \quad (7)$$

where $p_1 = \frac{1}{1+\mu/2}$ is the probability of a single event detection, and $p_2 = \frac{\mu/2}{1+\mu/2}$ is the probability to detect two events simultaneously. This equation allows us to calculate the $n''(E)$ considering the pileup effect if the value $\mu$ is known from the experiment.

We draw attention to one extremely important consequence. Let a function $f(E)$ be dependent on a set of $\{a_i\}$ parametrically. Let function $f^p(E)$ be derived from $f(E)$ via said transformation $n'(E) \to n''(E)$. It is clear that $f^p(E)$ is also parametrically dependent on the set $\{a_i\}$. Furthermore, due to the uniqueness of the transformation, any $f^p(E)$ with its parameters fixed has the unique $f(E)$ corresponding to it. Let now we know the spectrum observed in the presence of pileups, $n''(E)$. We are interested to exclude the pileup effect and find out the $n'(E)$. Suppose that we know from certain considerations that the spectrum $n''(E)$ is described by the function $f^p(E)$. Let's vary the parameters $\{a_i\}$ in such a way as to minimize the standard deviation $f^p(E)$ from $n''(E)$. Once this is done, the problem is solved: in fact, by finding the parameters minimizing the deviation $\{a_i^{min}\}$, we found not only the closest $f^p$ to $n''(E)$, but also the closest $f(E)$ to $n'(E)$. As such, the solution is

$$n'(E) = f(E)|_{\{a_i\}=\{a_i^{min}\}} .$$

Thus, we obtained the technique for the pileup effect exclusion, at least for the practically important special cases.

## 4. Results

To test the procedures developed we choose the experimental data obtained (main pulse: 60 mJ @ 800 nm 50 fs; prepulse: 160 mJ @ 1064 nm, 10 ns) (see [28] for more details on the experiment). It was accumulated from 2000 laser shots with the counting rate $\mu = 0.3$. The Geant4 model corresponds to the experimental arrangement exactly.

The following set of energies $\{E_{ei}\}$ was chosen: { 100, 200, 300, 400, 500, 600, 700, 800, 900, 1000, 1100, 1200, 1300, 1400, 1500, 1600, 1700, 1800, 1900, 2000, 2100, 2200, 2300, 2400, 2500, 2600, 2700, 2800, 2900, 3000, 3200, 3500, 3600, 3800, 4000, 4200, 4500, 4600, 4800, 5000, 5500, 6000, 6500, 7000, 7500, 8000, 8500, 9000, 9500, 10000 } keV . It took about 350 hours to calculate $\{n_i'(E)\}$ from this set using the server based on the Intel Xeon X5675 processor.

The set $\{E_{ie}\}$ and computation time needed at each $E_{ie}$ are presented in Fig.5. Note that the linear dependence in Figure 4 (b) means that the time required for a fairly accurate spectrum calculation with maximum energy $E_{e\,max}$ will grow linearly with $E_{e\,max}$.

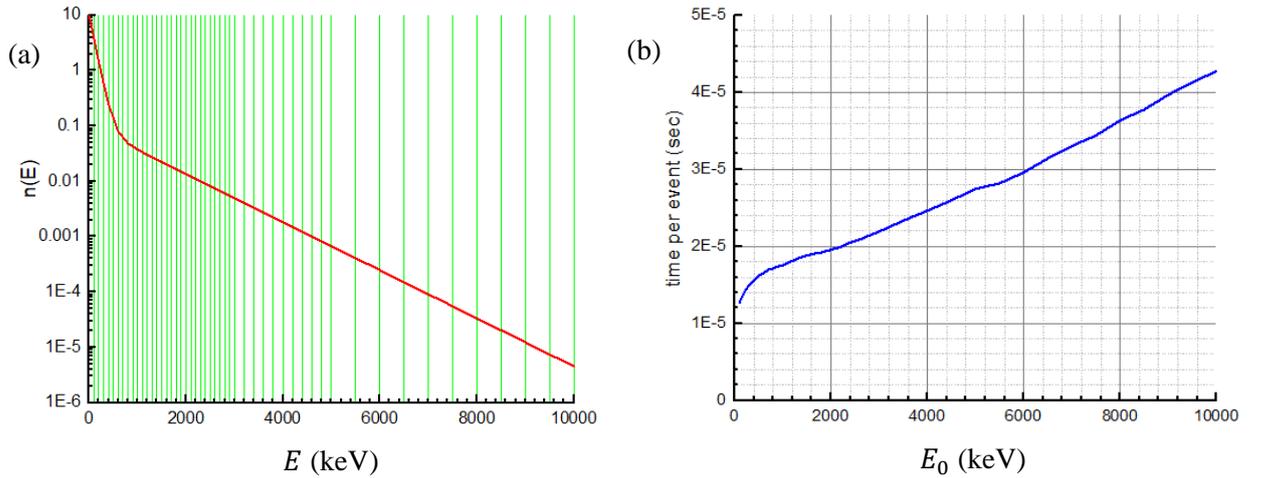

**Figure 4.** Graphical representation of the set $\{E_{ei}\}$ – mono energies (green vertical lines) and the expected envelope of the initial electron spectrum (red) (a) and dependence of the calculation time for one initial electron on its energy $E_e$. (b)

The set $\{n_i'(E)\}$ obtained for each $E_{ie}$ is presented in Figure 5. Then, the spectrum $n'(E) = \sum_{i=1}^{M} a_i n_i'(E)$ was calculated as an approximation to the spectrum for different $T$. Finally, each spectrum $n'(E)$ was fitted with the function $f(E, T') = \frac{1}{T} e^{-E/T'}$, yielding the $T'$ for each $T$. This way, function $T'(T)$ was obtained (see Figure 10 a from the Appendix 2). Note that this function depends only on the experimental geometry.

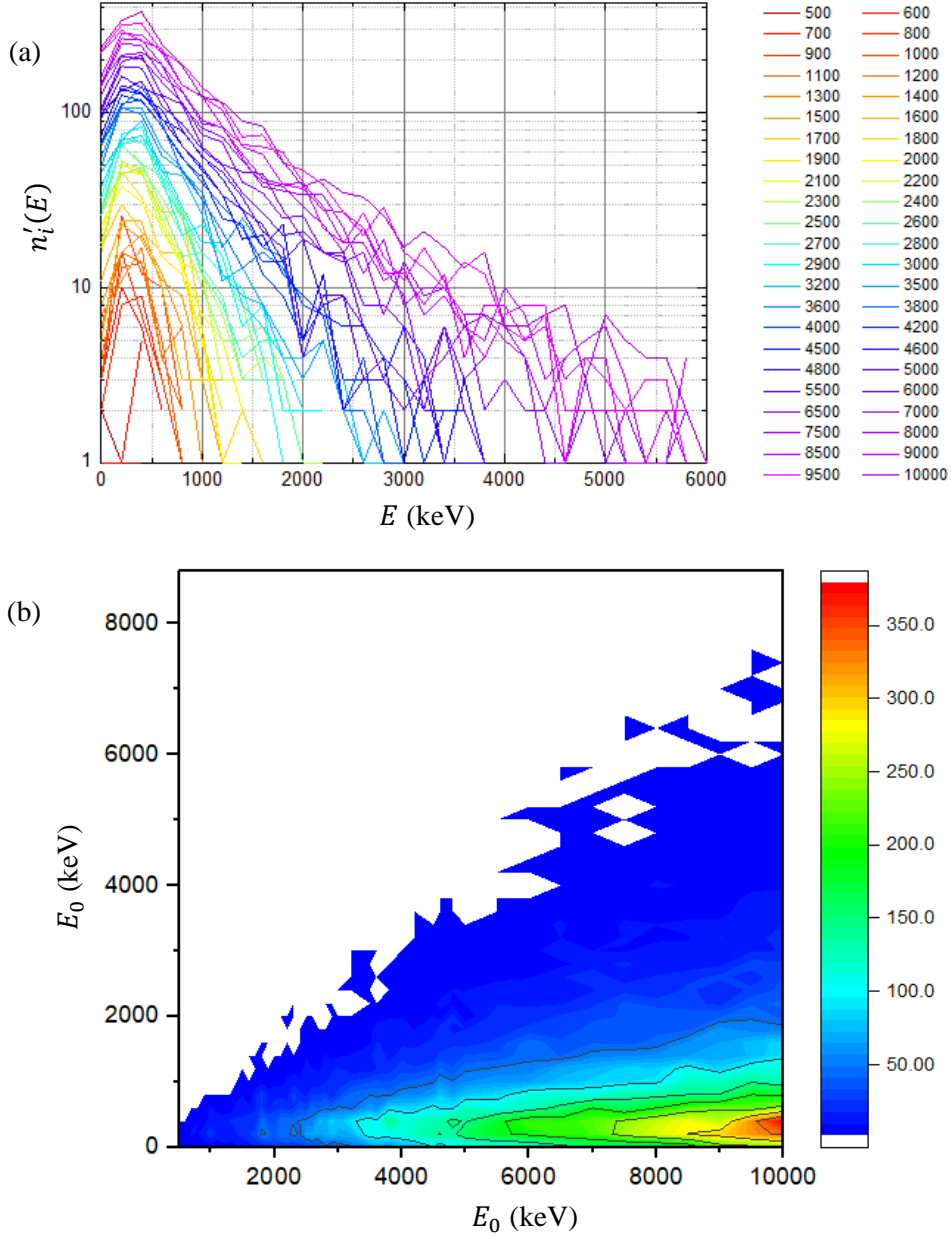

**Figure 5.** The $\{n'_i(E)\}$ set itself (a). Colours denote the $E_i$ (keV) the lines correspond to; Same, but in the equivalent form of a continuous two-dimensional function $p(E, E_{ye})$ – the kernel of the linear transformation $n(E_e) \to n'(E)$ (b).

To construct the final spectrum, we need to account for the pile-up effect. Assuming the spectrum $n'(E)$ to be bi-exponential, and knowing the counting rate $\mu$ and value of $p_2$, we fit the experimentally-acquired spectrum (which essentially is the $n''(E)$ spectrum) with the function ((34) (see Appendix 3) – a result of performing a pileup-transformation on said $n'$.

Our fitting yielded $T'_1 = 112 \pm 5$ keV and $T'_2 = 727 \pm 21 \; keV$ for the experimental data presented in Figure 6. Using the $T'(T)$ calculated earlier, we determine that these $n'$ parameters correspond to the $T_1 = 127 \pm 5 \; keV$ and $T_2 = 1990 \pm 100 \; keV$ for the initial bi-exponential electron spectrum $n(E_e)$ with $A/B = (1.3 \pm 0.2)\text{E}4$. All the functions mentioned above are shown in Figure 6. It should be noted that neglecting the pileup effect the fitting procedure yields $T'_1 = 132 \pm 6 \; keV$ and $T'_2 = 732 \pm 24 \; keV$, or $T_1 = 150 \pm 7 \; keV$ and $T_2 = 2050 \pm 120 \; keV$, i.e. overestimation of the spectrum temperatures occurs even for the $\mu = 0.3$.

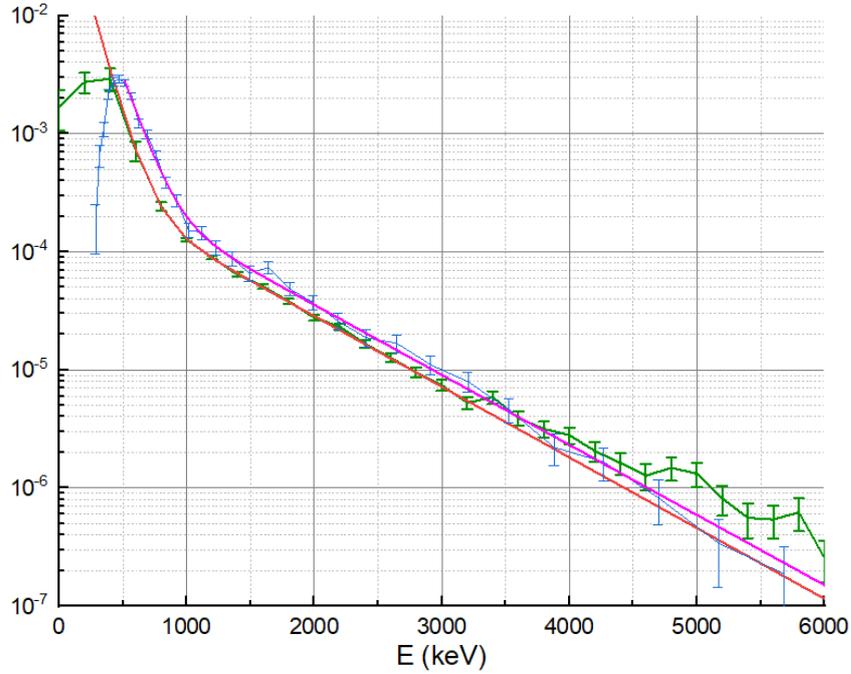

**Figure 6.** Experimentally observed spectrum ($n''(E)$) (blue), its approximation accounting for the pileup effect ($F_{BEP}$) (magenta), pileup-reverted experimental spectrum ($n'(E)$) (orange), $n_{appr}'(E)$ spectrum derived from calculations (green).

To double check our results, we used the estimated parameter set {$T_1$, $T_2$, A/B} to construct the pileup-free spectrum $n_{appr}'(E)$ and compare it with the experimental spectrum cleared from the pileup effect. This comparison is also shown in Figure 6. It is clear, that up to 3200 keV (which is more than $4T_2$) there is a good agreement between the two spectra. Seems to get even better coincidence the third exponential term must be introduced into consideration.

## 5. Conclusions

The experimentally measured gamma spectrum from the laser solid interaction at a high, relativistic intensity is often used to get information on the fast electron population in plasma. Direct modelling and fitting of such a spectrum using Monte-Carlo approach (for example, using GEANT 4 package) and accounting for the pileup events demands huge computational time that increases exponentially with increase in the energy range of the gamma spectrum under consideration. We suggested and backed novel approach reducing CPU time consumption significantly, making fitting procedure almost real-time.

Here first we computed and tabulated response of the experimental setup under consideration for a set of monoenergetic electron "events" acquiring high enough statistics within the wide range of gamma energies. This needs a lot of CPU time, but could be done once for the given setup. Next, we presented the initial guess for the electron spectrum with the sum of monoenergetic peaks yielding weight of each peak. Then we calculated the related gamma spectrum using the pre-tabulated data by summing spectra from each monoenergetic peak with their weights. To account for the pileup events, we developed analytical approach allowing to convert the calculated gamma spectrum into the one with the given probability of pileups. Finally, we compared thus converted spectrum to the experimental one. Further one needs to find the appropriate weights only, that can be implemented in the very robust and efficient code.

This algorithm was checked with the real gamma spectrum (within 0.5-4 MeV) obtained at the counting rate of $\mu$ =0.3 that corresponds to the essential impact from the pileup events. We successfully reconstructed the spectrum fitting it with bi=exponential function and got estimates on the slope of both exponents ($T_1 = 127 \pm 5\ keV$ and $T_2 = 1990 \pm 100\ keV$) and on the ratio of their amplitudes $A/B = 13000 \pm 2000$.

The approach developed can be used for different energy ranges and with different fitting functions for the spectrum of initial electrons. It is also applicable for similar tasks with other than gammas secondary particles.

**6. Acknowledgements**

Authors acknowledge helpful discussions with S.Shulyapov and D.Gorlova on different topics related to the paper.

This study was supported by the «Basis» Foundation (Grant #20-2-10-11-1).

This work was supported by RSF Grant #22-79-10087.

## Appendix 1. Derivation of the error of approximation of a continuous function with a discrete one

Let a continuous function $n(E)$ be approximated by a finite sum of discontinuous ones: segments of the approximate spectrum in the intervals between the chosen $E_i$ are considered only in the sense that the value of the coefficient $a_i$ is proportional to the proportion of this segment in the entire spectrum. It is obvious that the error will be the smaller, the closer the neighboring $E_i$ are located, and, accordingly, the more accurate the approximation of the energy of all particles from the interval $(E_{i-1}, E_i]$ by the value of $E_i$.

Let us estimate the value of this error. Let there be two spectra of initial particles: $n_C(E)$ (*Continuous*) and $n_D(E)$ (*Discrete*). We need to find the difference between the corresponding detected spectra $n'_C(E)$ and $n'_D(E)$ in an arbitrary point $E$, provided that:

$$n_D(E) = \sum_{i=1}^{\infty} \left( \delta(E - E_i) \cdot \int_{E_{i-1}}^{E_i} n_C(x)\, dx \right). \tag{8}$$

The linear transformation $n(E) \to n'(E)$ can be represented as:

$$n'(E) = \int_0^{\infty} p(E_e, E)\, n(E_e)\, dE_e, \tag{9}$$

where $E_0$— the energy of an initial electron. Expanding it for $n'_D(E)$:

$$\begin{aligned} n'_D(E) &= \int_0^{\infty} p(E_0, E) \sum_{i=1}^{\infty} \left( \delta(E_0 - E_i) \int_{E_{i-1}}^{E_i} n_C(x)\, dx \right) dE_0 = \\ &= \sum_{i=1}^{\infty} \left( \int_0^{\infty} p(E_0, E)\, \delta(E_0 - E_i)\, dE_0 \int_{E_{i-1}}^{E_i} n_C(x)\, dx \right) = \\ &= \sum_{i=1}^{\infty} \left( p(E_i, E) \int_{E_{i-1}}^{E_i} n_C(x)\, dx \right). \end{aligned} \tag{10}$$

Next, we perform transformations with $n'_C(E)$: first, we represent the integral along the semiaxis as the sum of integrals over the segments $[E_{i-1}, E_i]$, and then we use the Taylor expansion (keeping the first two terms of the series):

$$\begin{aligned} n'_C(E) &= \int_0^{\infty} p(E_0, E)\, n_C(E_0)\, dE_0 = \sum_{i=1}^{\infty} \int_{E_{i-1}}^{E_i} p(E_0, E)\, n_C(E_0)\, dE_0 \cong \\ &\cong \sum_{i=1}^{\infty} \int_{E_{i-1}}^{E_i} \left( p(E_i, E) + \left.\frac{\partial p(E_0, E)}{\partial E_0}\right|_{E_0 = E_i} (E_i - E_0) \right) n_C(E_0)\, dE_0 = \\ &= \sum_{i=1}^{\infty} p(E_i, E) \int_{E_{i-1}}^{E_i} n_C(E_0)\, dE_0 + \sum_{i=1}^{\infty} \int_{E_{i-1}}^{E_i} \left.\frac{\partial p(E_0, E)}{\partial E_0}\right|_{E_0 = E_i} \cdot (E_i - E_0)\, n_C(E_0)\, dE_0. \end{aligned} \tag{11}$$

Now it is easy to write the difference between the values:

$$\Delta n'_{DC}(E) = n'_D(E) - n'_C(E) = \sum_{i=1}^{\infty} \left.\frac{\partial p(E_0, E)}{\partial E_0}\right|_{E_0 = E_i} \cdot \int_{E_{i-1}}^{E_i} (E_i - E_0)\, n_C(E_0)\, dE_0. \tag{12}$$

Let's perform a Taylor expansion: $n_C(E_0) \cong n_C(E_i) + \frac{\partial n_C(E)}{\partial E}(E_i) \cdot (E_0 - E_i) + \cdots$. In order for this expansion to be able to discard terms beyond the $0^{th}$, the following must hold:

$$\left|\left(\frac{\partial n_C(E)}{\partial E} \cdot (E_i - E_{i-1})\right) \Big/ n_C(E)\right| \ll 1 \tag{13}$$

For our special case of $n_C(E) = \frac{1}{T} e^{-\frac{E}{T}}$ this is equivalent to: $\frac{(E_i - E_{i-1})}{T} \ll 1$. Then, if this condition holds true:

$$\int_{E_{i-1}}^{E_i} (E_i - E_0) n_C(E_0) \, dE_0 \cong n_C(E_i) \int_{E_{i-1}}^{E_i} (E_i - E_0) \, dE_0 == n_C(E_i) \cdot \frac{1}{2}(E_i - E_{i-1})^2 \tag{14}$$

And the error, associated with the discreteness of the function approximated, can be written as:

$$\Delta n'_{CD}(E) \cong \frac{1}{2} \sum_{i=1}^{\infty} \left.\frac{\partial p(E_0, E)}{\partial E_0}\right|_{E_0 = E_i} \cdot n_C(E_i) \cdot (E_i - E_{i-1})^2 \tag{15}$$

This shows that if the difference $E_i - E_{i-1}$ does not depend on $i$, and the partial derivative $\frac{\partial p(E_0, E)}{\partial E_0}$ changes slightly with increasing $E_0$, then the error increases quadratically with increasing distance between the approximating delta peaks.

If the condition $(E_i - E_{i-1})/T \ll 1$ is not met, then terms of larger orders will appear in this dependence – cubic, fourth degree, etc.

## Appendix 2. Derivation of the error of approximation a function with an infinite domain of definition with a function with a finite domain of definition

Let function $y(E_e)$ be the true initial electron spectrum and have $[0, \infty)$ as its definition domain, and $f(E_e) = \sum_{i=1}^{M} a_i \delta(E_e - E_{ei})$, $a_i = \int_{E_{ei-1}}^{E_{ei}} y(E_e) dE_e$ – its approximation.

The set $\{E_i\}$ is finite, therefore, it contains the largest energy $E_M$. The question arises: what value should this $E_M$ be in order for the difference between $f(E_e)$ от $y(E_e)$ to be sufficiently small in some interval of interest to us $E_e \in [E_{e1}, E_{e2}]$? To answer, it suffices to consider the influence of the initial particles on the ranges of the detected spectrum with upper boundaries lower than the energy the particles have.

First, consider the partition of the spectrum of initial particles $n(E_e)$ into two functions: $n(E_e) = f_A(E_e) + f_B(E_e)$, where $f_A \neq 0$ exclusively in $E_e \in [0, E_{e\,thr})$, and $f_B \neq 0$ exclusively in $E_e \in [E_{e\,thr}, \infty)$. Also, let $n'_A(E)$ and $n'_B(E)$ be the spectra detected in cases of $n(E_e) = f_A(E_e)$ and $n(E_e) = f_B(E_e)$, respectively. Using these notations, we define the quantity $\Delta$:

$$\Delta(E) = \frac{n'_B(E)}{n'_A(E) + n'_B(E)}. \tag{16}$$

It shows what fraction of the value $n'$ at point $E$ is lost if $f_B$ is discarded from the spectrum of initial particles $n(E_e)$, i.e. the part to the right of $E_{e\,thr}$. If in the interval $[E_{e1}, E_{e2}]$ the value of $\Delta$ is sufficiently small, then the contribution of the discarded part $n'_B(E)$ can be neglected, which means that the chosen $E_{e\,thr}$ is high enough.

In case of the discrete representation of $n(E_e)$, the previously introduced value $\Delta$ will be written as:

$$\Delta = \frac{\sum_{i=M+1}^{\infty} a_i n'_i(E)}{\sum_{i=1}^{\infty} a_i n'_i(E)} = \frac{1}{\frac{\sum_{i=1}^{M} a_i n'_i(E)}{\sum_{i=M+1}^{\infty} a_i n'_i(E)} + 1}. \tag{17}$$

In order to find the $M$ providing sufficiently low $\Delta$, it is necessary to make some assumptions about the dependence of $n'_i(E)$ on $i$ with $E$ fixed.

To do that, we must first represent the $\{n'_i(E)\}$ set in the form of a function $p(E, E_e)$, determining the conditional probability of an event of energy $E$ to be detected, given the initial electron energy was $E_e$. This two-dimensional function is, essentially, an interpolation between individual $n'_i(E)$, as those are the probabilities of an event of energy $E$, given the initial electron energy was $E_e = E_{ei}$ from the finite set $\{E_{ei}\}$. Resulting interpolated function is presented in **Figure 5**. Figure 7 shows the function $p(E, E_e)$ for some values of $E$. It is obvious that the dependence of $p(E, E_e)$ on $E_e$ is close to linear and can be fitted with $\alpha(E)E_e + \beta(E)$. Figure 8 shows the dependence of the coefficients $\alpha(E)$ and $\beta(E)$. The slope $\alpha$ decreases with increasing $E$. Given the set $\{E_{ei}\}$ is evenly spaced, that is, the distance between the peaks $E_{ei} - E_{e\,i-1}$ does not depend on $i$, the dependence of $n'_i(E)$ on $i$ at a fixed $E$ is also linear.

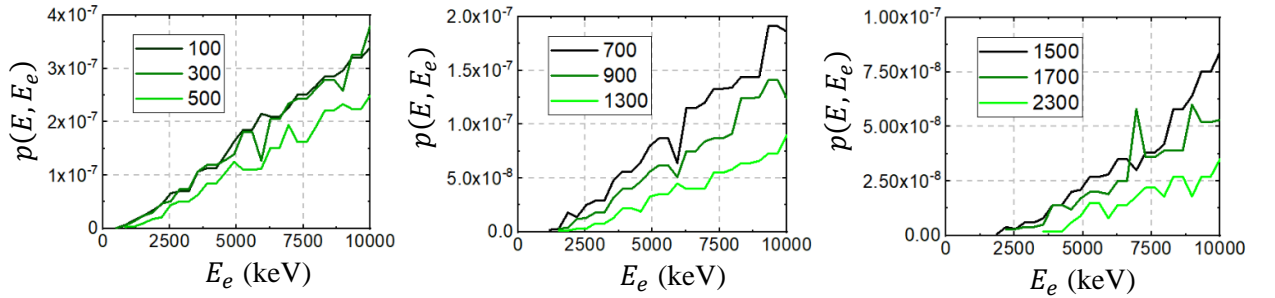

**Figure 7**. Dependence of $p(E, E_e)$ on $E_e$ at various constant $E$ (value indicated by color, in keV).

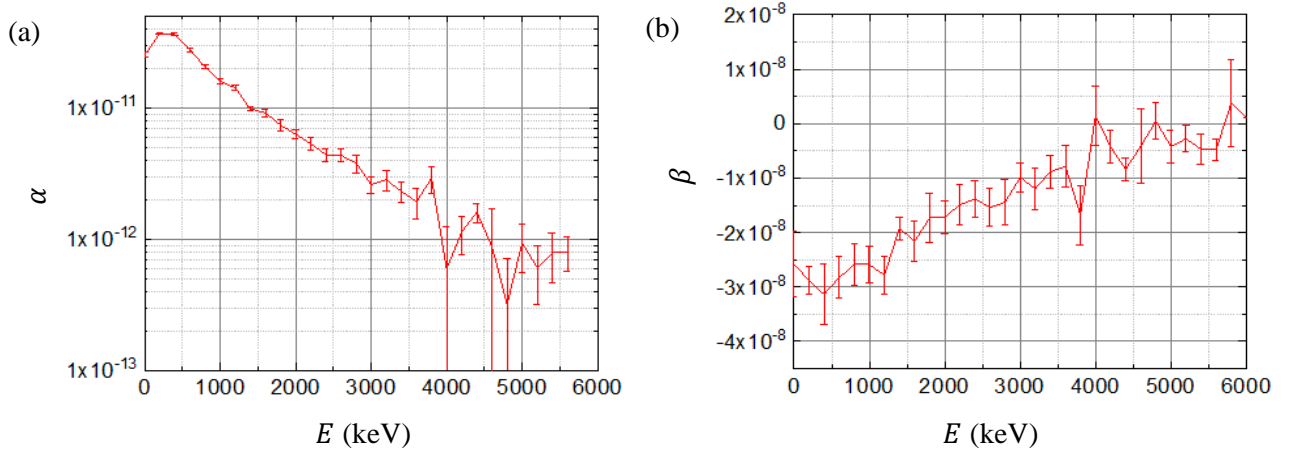

**Figure 8.** Dependence of $\alpha$ (a) and $\beta$ (b) on $E$ (keV) in fitting the $p(E, E_e)$ with the $g(E_e) = \alpha E_e + \beta$.

In the case of an exponential spectrum of initial electrons:

$$n(E_e) = y(E_e) = \frac{1}{T}e^{-E_e/T}, \qquad (18)$$

$$a_i = \int_{E_{ei-1}}^{E_{ei}} y(E_e)dE_e = \int_{E_{ei-1}}^{E_{ei}} \frac{1}{T}e^{-E_e/T}dE_e = e^{-\frac{E_{ei}}{T}}\left(e^{-\frac{E_{ei}-E_{ei-1}}{T}} - 1\right) = e^{-\frac{E_{ei}}{T}} * C. \qquad (19)$$

The terms $a_i n'_i(E)$ from the infinite sums in (17) can be written as:

$$a_i n'_i(E) = e^{-\frac{E_{ei}}{T}} * C * (\alpha E_e + \beta), \qquad (20)$$

which has the asymptote of $i * e^{-i}$. The dependence of $\Delta$ on the number of peaks $M$ is shown in the **Figure 9**. It can be seen that this error decreases rather quickly, almost exponentially – a decrease in $\Delta(x)$ by almost 2 times with the addition of one more peak to the set $\{E_i\}$.

From the foregoing, we can conclude that this happens at all points of the spectrum simultaneously. Here it is necessary, to note that the linear dependence $n'_i(E) = i * \Delta E$ is approximate, and its slope varies depending on $E$. We also recall that for any $n'_i(E)$ there are no non-zero points above $E_{ei}$, due to the law of conservation of energy. Therefore, for the point $E_e \in (E_{ej}, E_{ej+1}]$ the value of $M$ will be $M - j$, because the spectra $n'_1 \ldots n'_j$, obtained from the first $j$ delta peaks of the sum, have null contribution to the sum at that point.

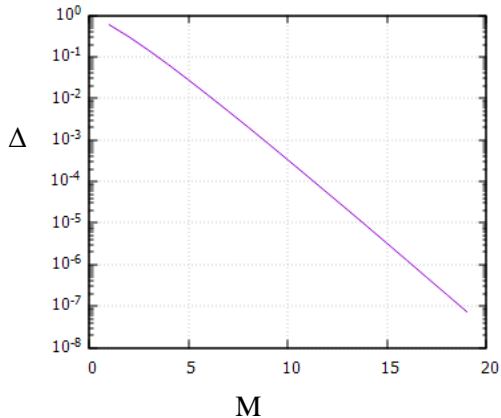

**Figure 9.** The dependence of $\Delta$ on $M$.

Thus, the value of this error at a given point is affected by the number of delta peaks to the right of it in the set $\{E_{ei}\}$. The dependence $T'(T)$ for various end energies of a series of delta peaks from $\{E_{ei}\}$ is shown in Figure 10a.

Let's explore this dependence. Let the lower index at $T'$ denote the termination energy ($E_{ei}$) of the sum $n'(E) = \sum_i (a_i * n'_i(E))$. For example, $T'_{E_{max}}$ means that the said sum was calculated up to $E_{max}$. It is clear that the most accurate dependence $T'_E(T)$ known to us is the one that uses all the available $n'_i(E)$ – namely, $T_{10MeV}'(T)$. The location of the branching points – the points at which $T'(T)$ deviate from the $T_{10MeV}'(T)$ by more than 1%, – is presented in Figure 10b.

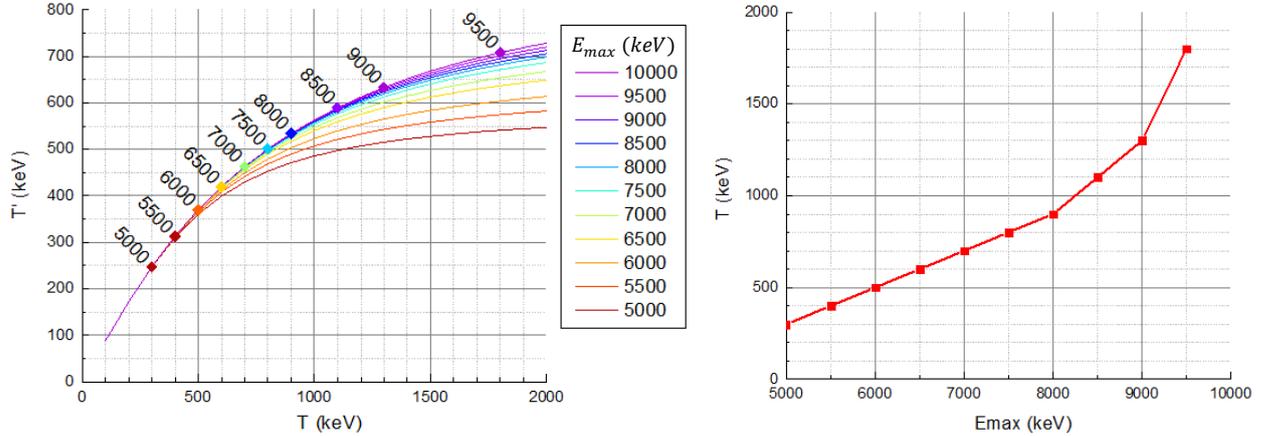

**Figure 10.** The dependence $T'(T)$ for the various termination energies of the sum (color-coded, in keV) (dots denote branching points) (a) and the dependence of the branching energy on the sum's termination energy $E_{max}$ (b).

Thus, we can assume that the set of energies $\{E_{ei}\}$ used in our example (see **Figure 5**) is sufficient for the values of $T$ up to 2 MeV.

**Appendix 3. Derivation of an analytical expression for the transformation of the detected spectrum due to the pileup effect**

**1. Formulation**

It is known that, on average, there are $\mu$ detector events per one initial laser pulse. In addition, it is known that if the pileup is absent, i.e. one detector event is always triggered by one particle, or, equivalently, $\mu \ll 1$, the spectrum measured by the detector would be equal to $P(E)$.

What spectrum does the detector measure if the detector event is triggered by simultaneously $\mu$ particles on average?

**2. Case of two particles (first order approximation).**

We assume that the particles arrive at the detector independently from each other. Then the probability of registering $k$ particles will be determined by the Poisson distribution:

$$P(k) = \frac{\mu^k}{k!} e^{-\mu} \qquad (21)$$

We assume that the probability to detect more than two particles simultaneously can be neglected. The probability that a single particle is detected reads as

$$p_1 = P(k=1 | k=1,2) = \frac{P(k=1)}{P(k=1) + P(k=2)} = \frac{1}{1 + \frac{\mu}{2}}. \qquad (22)$$

The probability that two particles are detected simultaneously is

$$p_2 = P(k=2 | k=1,2) = \frac{P(k=1)}{P(k=1) + P(k=2)} = \frac{\frac{\mu}{2}}{1 + \frac{\mu}{2}} \qquad (23)$$

These conditional probabilities determine what is detected by the detector, and they will be of importance in further consideration, as the measured spectrum consists only of those cases when a detector event occurred.

**2.1. Formulation**

Let us reformulate this problem for the case when the contribution of events of registration of more than two particles simultaneously can be neglected. It is clear that this simplification can be considered legitimate provided $\mu + \mu^2 \gg \sum_{k=3}^{\infty} \mu^k$.

Let $\alpha, E_1, E_2$ be independent random variables. Let $E_1$ and $E_2$ have probability density distributions set, respectively, by the functions $P_{E1}$ and $P_{E2}$. Let the value of $\alpha$ be 1 with probability $p_2$, and 0 with probability $p_1 = 1 - p_2$.

Then, random variable $\eta = E_1 + \alpha E_2$ corresponds to the energy detected by the detector in one event, and its probability distribution $P_\eta$ defines the spectrum observed by the detector. Our task is to find $P_\eta$, knowing $p_1, p_2, P_{E1}$ and $P_{E2}$.

**2.2. Solution**

Let's find the distribution function $F_\eta(y)$ for $\eta$. Let the $x_1$ be a random value of $E_1$, $x_2$ be a random value of $E_2$, and $a$ be that of $\alpha$. Then:

$$\begin{aligned} F_\eta(y) = P(\eta < y) &= P(x_1 < y | a = 0) + P(x_1 + x_2 < y | a = 1) = \\ &= P(x_1 < y) p_1 + P(x_1 + x_2 < y) p_2 \end{aligned} \qquad (24)$$

Now we can write an expression for $P_\eta(y)$:

$$P_\eta(y) = \frac{\partial}{\partial y} F_\eta(y) = p_1 \frac{\partial}{\partial y} P(x_1 < y) + p_2 \frac{\partial}{\partial y} P(x_1 + x_2 < y) \quad (25)$$

The first term is trivially expanded:

$$P(x_1 < y) = \int_0^y P_{E1}(x_1)\, dx_1 \;; \quad (26)$$

$$\frac{\partial}{\partial y} P(x_1 < y) = \frac{\partial}{\partial y} \int_0^y P_{E1}(x_1)\, dx_1 = P_{E1}(y) \quad (27)$$

Let's expand the second term.

$$P(x_1 + x_2 < y) = \iint_{x_1+x_2<y} P_{E1}(x_1) P_{E2}(x_2) * dx_1 * dx_2 =$$
$$= \int_0^y P_{E2}(x_2) \left( \int_0^{y-x_2} P_{E1}(x_1) * dx_1 \right) * dx_2 == \int_0^y P_{E2}(x_2) * (F_{E1}(y - x_2) - F_{E1}(0)) * dx_2 \quad (28)$$

For further transformation we need the following *lemma*.

*Lemma.*
Claim:

$$\frac{\partial}{\partial y} \left( \int_0^{b(y)} f(x,y)\, dx \right) = b'(y) * f(b(y), y) + \int_0^{b(y)} \frac{\partial}{\partial y} f(x,y)\, dx \quad (29)$$

Proof:

$$\frac{\partial}{\partial y} \left( \int_0^{b(y)} f(x,y)\, dx \right) = \lim_{\Delta \to 0} \frac{1}{\Delta} \left( \int_0^{b(y+\Delta)} f(x, y+\Delta)\, dx - \int_0^{b(y)} f(x,y)\, dx \right) \quad (30)$$

we use the Taylor series expansion in $\Delta$ in the vicinity of $y$. Then:

$$\frac{\partial}{\partial y} \left( \int_0^{b(y)} f(x,y)\, dx \right) =$$
$$= \lim_{\Delta \to 0} \frac{1}{\Delta} \left( \int_0^{b(y+\Delta)} \left( f(x,y) + \frac{d}{dy} f(x,y) * \Delta + \frac{d^2}{dy^2} f(x,y) * \frac{\Delta^2}{2} + \cdots \right) dx - \int_0^{b(y)} f(x,y)\, dx \right) =$$
$$= \lim_{\Delta \to 0} \frac{1}{\Delta} \left( \int_0^{b(y+\Delta)} f(x,y) * dx - \int_0^{b(y)} f(x,y) * dx + \right.$$
$$\left. + \int_0^{b(y+\Delta)} \left( \frac{d}{dy} f(x,y) * \Delta + \frac{1}{2} \frac{d^2}{dy^2} f(x,y) * \Delta^2 + \cdots \right) dx \right) = \quad (31)$$

$$= \lim_{\Delta \to 0} \frac{\int_0^{b(y+\Delta)} f(x,y) * dx - \int_0^{b(y)} f(x,y) * dx}{\Delta} +$$

$$+ \lim_{\Delta \to 0} \int_0^{b(y+\Delta)} \left( \frac{d}{dy} f(x,y) + \frac{1}{2} \frac{d^2}{dy^2} f(x,y) * \Delta + \cdots \right) dx.$$

Let $F(x_1, y) = \int_0^{x_1} f(x,y) * dx$. Then:

$$\frac{\partial}{\partial y} \left( \int_0^{b(y)} f(x,y) \, dx \right) =$$

$$= \lim_{\Delta \to 0} \frac{F(b(y+\Delta), y) - F(b(y), y)}{\Delta} + \int_0^{b(y)} \frac{d}{dy} f(x,y) * dx =$$

$$= \frac{\partial}{\partial y_1} F(b(y_1), y) \bigg|_{y_1 = y} + \int_0^{b(y)} \frac{\partial}{\partial y} f(x,y) \, dx = \quad (32)$$

$$= b'(y) * \left( \frac{\partial}{\partial y_1} F(y_1, y) \right) \bigg|_{y_1 = b(y)} + \int_0^{b(y)} \frac{\partial}{\partial y} f(x,y) \, dx =$$

$$= b'(y) * f(b(y), y) + \int_0^{b(y)} \frac{\partial}{\partial y} f(x,y) \, dx.$$

The claim is proven.

*Continuation.*
Now it is easy to calculate:

$$\frac{\partial}{\partial y} P(x_1 + x_2 < y) = \frac{\partial}{\partial y} \int_0^y P_{E2}(x_2) * \left( F_{E1}(y - x_2) - F_{E1}(0) \right) * dx_2 =$$

$$= P_{E2}(y) * \left( F_{E1}(y - y) - F_{E1}(0) \right) + \int_0^y P_{E2}(x_2) * \frac{\partial}{\partial y} F_{E1}(y - x_2) * dx_2 = \quad (33)$$

$$= \int_0^y P_{E2}(x) * P_{E1}(y - x) * dx$$

The desired probability distribution density is:

$$P_\eta(y) = p_1 * P_{E1}(y) + p_2 * \int_0^y P_{E2}(x) * P_{E1}(y - x) * dx := p_1 * P_{E1}(y) + p_2 * P_{E1E2}(y) \quad (34)$$

The meaning of this expression is obvious: a two-particle term is superimposed on the initial one-particle distribution. It is also easy to see that the normalization to unity is preserved.

### 2.3. Important special cases

*Single exponential spectrum*
Let: $P_{E1}(x) = P_{E2}(x) = \frac{1}{T} e^{-\frac{x}{T}}$. Then:

$$P_{E1E2}(y) = \int_0^y \frac{1}{T} e^{-\frac{x}{T}} * \frac{1}{T} e^{-\frac{y-x}{T}} * dx = \frac{1}{T^2} \int_0^y e^{-\frac{y}{T}} * dx = \frac{1}{T^2} e^{-\frac{y}{T}} * y, \tag{35}$$

$$P_\eta(y) = p_1 * \frac{1}{T} e^{-\frac{x}{T}} + p_2 * \frac{1}{T^2} e^{-\frac{y}{T}} * y, \tag{36}$$

or, equivalently,

$$P_\eta(y) = \frac{1}{T^2} e^{-\frac{y}{T}} * \left(p_1 + p_2 * \frac{y}{T}\right). \tag{37}$$

*Two-component spectrum*
Let $P_{E1}(x) = P_{E2}(x) = P_1(x) + P_2(x)$. Then:

$$P_{E1E2}(y) = \int_0^y (P_1(x) + P_2(x)) * (P_1(y-x) + P_2(y-x)) * dx =$$

$$= \int_0^y P_1(x) P_1(y-x) dx + \int_0^y P_1(x) P_2(y-x) dx + \int_0^y P_2(x) P_1(y-x) dx \tag{38}$$

$$+ \int_0^y P_2(x) P_2(y-x) dx := P_{11}(y) + P_{12}(y) + P_{21}(y) + P_{22}(y).$$

It is quite easy to show that $P_{12}(y) = P_{21}(y)$. Knowing this, one can write:

$$P_{E1E2}(y) = P_{11}(y) + 2 * P_{12}(y) + P_{22}(y). \tag{39}$$

*Bi-exponential Spectrum*
Let: $P_{E1}(x) = P_{E2}(x) = A_1 e^{-\frac{x}{T_1}} + A_2 e^{-\frac{x}{T_2}}$, where $T_1 \neq T_2$.
Then:

$$P_{21} = P_{12} = \int_0^y P_1(x) P_2(y-x) dx = \int_0^y A_1 A_2 e^{-\frac{x}{T_1} - \frac{y}{T_2} + \frac{x}{T_2}} dx = A_1 A_2 e^{-\frac{y}{T_2}} \int_0^y e^{-\frac{T_2-T_1}{T_1 T_2}x} dx =$$

$$= A_1 A_2 \frac{T_1 T_2}{T_2 - T_1} e^{-\frac{y}{T_2}} \int_0^{y/T_{eff}} e^{-x} dx = A_1 A_2 \frac{T_1 T_2}{T_2 - T_1} e^{-\frac{y}{T_2}} \left(1 - e^{-\frac{T_2-T_1}{T_1 T_2}y}\right) \tag{40}$$

$$= A_1 A_2 \frac{T_1 T_2}{T_2 - T_1} \left(e^{-\frac{y}{T_2}} - e^{-\frac{y}{T_1}}\right);$$

$$P_{11} = \int_0^y A_1^2 e^{-\frac{x}{T_1} - \frac{y}{T_1} + \frac{x}{T_1}} * dx = A_1^2 \int_0^y e^{-\frac{y}{T_1}} * dx = A_1^2 * y * e^{-\frac{y}{T_1}}; \tag{41}$$

$$P_{22} = A_2^2 * y * e^{-\frac{y}{T_2}}. \tag{42}$$

In this case:

$$P_\eta(y) = p_1 * \left(A_1 e^{-\frac{y}{T_1}} + A_2 e^{-\frac{y}{T_2}}\right) + \tag{43}$$

– 20 –

$$+ p_2 * \left(A_1^2 y e^{-\frac{y}{T_1}} + 2 A_1 A_2 \frac{T_1 T_2}{T_2 - T_1}\left(e^{-\frac{y}{T_2}} - e^{-\frac{y}{T_1}}\right) + A_2^2 y e^{-\frac{y}{T_2}}\right)$$

Or, equivalently:

$$P_\eta(y) = \left(p_1 - p_2 * 2 A_2 \frac{T_1 T_2}{T_2 - T_1}\right) * A_1 e^{-\frac{y}{T_1}} + p_2 * A_1^2 * y e^{-\frac{y}{T_1}} +$$
$$+ \left(p_1 - p_2 * 2 A_1 \frac{T_1 T_2}{T_1 - T_2}\right) * A_2 e^{-\frac{y}{T_2}} + p_2 * A_2^2 * y e^{-\frac{y}{T_2}}.$$
(44)

For convenience, we denote this function as $F_{BEP}$ (*bi-exponential pileup*).

Let's perform some asymptotic estimates, assuming that $T_2 \gg T_1$:

$$\frac{P_{12}}{P_{11}} = \frac{A_2}{A_1} \frac{T_2 - T_1}{T_1 T_2} \frac{1}{y}\left(e^{\left(\frac{1}{T_1} - \frac{1}{T_2}\right)y} - 1\right) \sim \frac{e^{\frac{T_2 - T_1}{T_1 T_2} * y} - 1}{y} \to \begin{cases} 1 &, & y \to 0 \\ \infty &, & y \to \infty \end{cases} \quad (45)$$

$$\frac{P_{12}}{P_{22}} = \frac{A_1}{A_2} \frac{T_2 - T_1}{T_1 T_2} \frac{1}{y}\left(1 - e^{-\left(\frac{1}{T_1} - \frac{1}{T_2}\right)y}\right) \sim \frac{1 - e^{-\frac{T_2 - T_1}{T_1 T_2} * y}}{y} \to \begin{cases} 1 &, & y \to 0 \\ 0 &, & y \to \infty \end{cases} \quad (46)$$

Thus, none of the terms in (44) can be neglected.